\documentclass[10pt,twocolumn,amssym,amsmath,bm]{./IEEEtran_v15}
\usepackage{epsfig}

\makeatletter
\def\ifundefined{\@ifundefined}
\makeatother

\begin{document}

\title{\Large \textbf{
Real-time formalism for studying the nonlinear response 
of ``smart'' materials to an electric field}
}

    \author{\large J.~K.~Freericks\\
    \normalsize \textit{ Department of Physics}\\
    \normalsize \textit{ Georgetown University}\\
    \normalsize  \textit{Washington, DC 20057}\\
    \normalsize  \textit{Email: freericks@physics.georgetown.edu}
    \and
    \large V.~M.~Turkowski\\
    \normalsize \textit{Department of Physics}\\
    \normalsize \textit{Georgetown University}\\
    \normalsize \textit{Washington, DC 20057}\\
    \normalsize \textit{Email: turk@physics.georgetown.edu}
    \and
    \large V.~Zlati\'c\\
    \normalsize \textit{Institute of Physics}\\
    \normalsize \textit{Bijenicka c. 46, P. O. B. 304}\\
    \normalsize \textit{10000 Zagreb, Croatia}\\
    \normalsize \textit{Email: zlatic@ifs.hr}}

% The major version number of the class file will not
% be defined with the old IEEEtran.cls. So, we can use this fact
% to determine if we are running the old or the new class.
\ifundefined{IEEEtransversionmajor}{%
   % This block will be executed only if we are using the old
   % class. All we do is to make sure the V1.3 lengths and commands
   % actually exist so the code won't choke when it
   % doesn't find them.
   
   % This file doesn't need most of these definitions.
   % However, we'll provide them all in case somebody
   % wants to see what should be executed when compiling
   % a V1.3 or later IEEEtran.cls .tex file with a pre V1.3
   % IEEEtran.cls class file. In such a case, all you have to
   % do is copy this block to the start of your code. 
   % However, it won't fix any bugs in the old IEEEtran.cls!
   %
   % **** BACKWARD COMPATIBILITY CODE BLOCK START ****   
   \newlength{\IEEEilabelindent}
   \newlength{\IEEEilabelindentA}
   \newlength{\IEEEilabelindentB}
   \newlength{\IEEEelabelindent}
   \newlength{\IEEEdlabelindent}
   \newlength{\labelindent}
   \newlength{\IEEEiednormlabelsep}
   \newlength{\IEEEiedmathlabelsep}
   \newlength{\IEEEiedtopsep}

   \providecommand{\IEEElabelindentfactori}{1.0}
   \providecommand{\IEEElabelindentfactorii}{0.75}
   \providecommand{\IEEElabelindentfactoriii}{0.0}
   \providecommand{\IEEElabelindentfactoriv}{0.0}
   \providecommand{\IEEElabelindentfactorv}{0.0}
   \providecommand{\IEEElabelindentfactorvi}{0.0}
   \providecommand{\labelindentfactor}{1.0}
   
   \providecommand{\iedlistdecl}{\relax}
   \providecommand{\calcleftmargin}[1]{
                   \setlength{\leftmargin}{#1}
                   \addtolength{\leftmargin}{\labelwidth}
                   \addtolength{\leftmargin}{\labelsep}}
   \providecommand{\setlabelwidth}[1]{
                   \settowidth{\labelwidth}{#1}} 
   \providecommand{\usemathlabelsep}{\relax}
   \providecommand{\iedlabeljustifyl}{\relax}
   \providecommand{\iedlabeljustifyc}{\relax}
   \providecommand{\iedlabeljustifyr}{\relax}
 
   \newif\ifnocalcleftmargin
   \nocalcleftmarginfalse

   \newif\ifnolabelindentfactor
   \nolabelindentfactorfalse
   
   % in V1.4 of IEEEtran.cls
   \newif\ifcenterfigcaptions
   \centerfigcaptionsfalse
   
   % we need to provide the old IED environments
   % with a bogus optional argument
   \let\OLDitemize\itemize
   \let\OLDenumerate\enumerate
   \let\OLDdescription\description
   
   \renewcommand{\itemize}[1][\relax]{\OLDitemize}
   \renewcommand{\enumerate}[1][\relax]{\OLDenumerate}
   \renewcommand{\description}[1][\relax]{\OLDdescription}

   \providecommand{\pubid}[1]{\relax}
   \providecommand{\pubidadjcol}{\relax}
   \providecommand{\specialpapernotice}[1]{\relax}
   \providecommand{\overrideIEEEmargins}{\relax}
   
   % V1.1 change: use \let instead of \providecommand
   % This prevents LaTeX from hanging if the user ever
   % tried to redefine \PARstart in terms of \CMPARstart 
   \let\CMPARstart\PARstart 
   
   \let\OLDappendix\appendix
   \renewcommand{\appendix}[1][\relax]{\OLDappendix}
   
   \newif\ifuseRomanappendices
   \useRomanappendicestrue
   
   % V1.2 change: handle the optional biography environment argument
   % (the photo specifier) provided by IEEEtran V1.5 and later.
   % This is tricky because, under the new biography, the SECOND
   % argument is the non-optional one (the biography text).
   \let\OLDbiography\biography
   \let\OLDendbiography\endbiography
   \renewcommand{\biography}[2][\relax]{\OLDbiography{#2}}
   \renewcommand{\endbiography}{\OLDendbiography}
   % **** BACKWARD COMPATIBILITY CODE BLOCK END ****
   
   % alter the header to show we are using the older class
   \markboth{A Test for IEEEtran.cls--- {\tiny \bfseries
   [Running Older Class]}}{Shell: A Test for IEEEtran.cls}}{
   % END IF OLDER CLASS 
  
   % This block will be executed only if we are running 
   % the enhanced class
   % alter the header to show we are using the enhanced class
%  \markboth{IEEE--- {\tiny \bfseries
%  [Running Enhanced Class
%   V\IEEEtransversionmajor.\IEEEtransversionminor]}}%
   }
% end of conditional

% Uncomment this line to render the big first letter in
% Computer Modern font. 
%\renewcommand{\PARstart}[2]{\CMPARstart{#1}{#2}}
% V1.1 change: This would hang if using the older IEEEtran.cls
% in IEEEtest V1.0, it is OK now.
%
%
% Here's how you would do invited papers:
%\specialpapernotice{(Invited Paper)}
% If you are binding copies of work generated with IEEEtran.cls,
% you may want to try:
%\overrideIEEEmargins
% These commands work OK here even though they are not in the preamble.
%(I wanted to put them after the backward compatibility code)

\maketitle
     \thispagestyle{empty}
     \pagestyle{empty}

% here's how you get a publisher's ID mark with the new
% IEEEtran.cls.  If you want to use it, don't forget to
% also uncomment the \pubidadjcol command (which must be
% executed in the second text column) around line 434 below
%\pubid{0000--0000/00\$00.00~\copyright~2001 IEEE}

\begin{abstract}
The nonlinear response of a material to a large electric field
(steady or pulsed) often determines the ultimate performance of the
material for electronics applications.  The formalism for
understanding nonlinear effects in conventional semiconductors is well
understood.  The formalism is less well developed for so-called ``smart''
materials that are tuned to lie close to the metal-insulator transition.
Here we show that the nonlinear response of a strongly correlated
electronic material can be calculated with a massively parallel algorithm
by discretizing a continuous matrix operator on the Kadanoff-Baym contour
in real time.  We benchmark the technique by examining the solutions
when the field vanishes and comparing the results to exact results from
an equilibrium formalism.  We briefly discuss the numerical issues associated
with the case of a large electric field and present results that show how
the Bloch oscillations become damped as the scattering due to electron 
correlations increases.
\end{abstract}

% no need for this single page document
% you may have to move \pubidadjcol (if used) if
% these are enabled
%\listoffigures
%\listoftables
%\tableofcontents

\section{Introduction}
\PARstart{T}{he} 
problem of the response of materials used in electronics to large 
external fields
is important from both a theoretical and a practical point of view.
On the theoretical side, the basic ideas of nonequilibrium statistical
mechanics were developed over 40 years ago~\cite{kadanoff_baym,keldysh},
but the formalism has not been  applied to strongly correlated materials except
in approximate ways.  It is interesting to determine exact results for
electronic systems in an external field which can be used to benchmark these
approximate techniques. On the practical side, it is often 
the nonlinear behavior
of the material or device that determines the ultimate performance within
electronics.  For example, the nonlinear current-voltage characteristic of
a $p$-$n$ junction is critical for semiconductor-based switching and digital
logic, while the nonlinear current-voltage characteristic of a nonhysteretic
Josephson junction allows for digital logic based on rapid single flux
quantum (RSFQ) ideas~\cite{rsfq}. 
A ``smart'' material is a material that can have
its properties altered by changing an external system variable like pressure,
temperature, or a gate voltage.  The most common devices with tunability are
currently based on semiconductors or ferroelectrics, but there is increasing 
interest in
strongly correlated materials near the metal-insulator transition, because
they might allow for more tunability than their semiconducting counterparts.

The interest in large electric fields arises as the system dimensions 
shrink onto the nanoscale.  When a feature size is on the order of
$100$~nm, a potential difference of $1$~V produces an electric field of
$E\sim 10^{7}$~V/cm over the feature area. In addition, the military is 
interested
in the robustness of devices to large pulsed fields that can arise from
natural sources like lightning, or from man-made sources like
those employed in electronic warfare.
These high energy-density short-time pulsed fields may be difficult to 
filter out of a device and can cause the device to ``burn out''.

A ``smart material'' tuned to lie close to the metal-insulator transition
is called a strongly correlated material.  The name arises from the fact
that one needs to take into account the electron-electron repulsion
in determining how the material responds to external perturbations. In
conventional metals, insulator, and semiconductors, it is adequate to
ignore the mutual electron-electron repulsion, and treat all of the electrons
as independent, moving in an average field created by the other electrons.
This is the regime where band-theory holds.  But as the electron-electron
interactions are made stronger relative to the kinetic energy of the 
electrons, then the electron correlations need to be taken into account,
implying that one cannot treat the other electrons in an averaged way, but one
needs to take into account where the electrons are and how they move
as every other electron moves. This is the regime of strong electron 
correlations, and as the correlations are increased, many materials will
undergo a metal-insulator transition called the Mott-Hubbard transition.

The simplest model which takes into account strong electron-electron
correlations is the Falicov-Kimball model \cite{falicov_kimball}.
This model has two kinds of electrons, itinerant electrons and localized
electrons.  They interact by a Coulomb repulsion  when they both occupy
the same unit cell of the lattice. If the number of itinerant electrons plus
the number of localized electrons is equal to the number of lattice sites,
then the system will undergo a metal-insulator transition as the Coulomb
repulsion is increased. This model is not appropriate to describe many
real materials, but its simplicity allows for many exact results to be 
calculated which are vitally important for benchmarking purposes.

\section{\label{sec:level2} Formalism}

We consider the Falicov-Kimball (FK) model in the presence of
an external electric field  that is spatially uniform, 
but can be time-dependent, and can have an arbitrarily large amplitude.
The FK model has two kinds of electrons: itinerant electrons
with creation and annihilation operators $c_{i}^{\dagger}$ and $c_{i}$ for 
conduction electrons at site $i$
and localized electrons with the corresponding operators
$f_{i}^{\dagger}$ and $f_{i}$.
The FK Hamiltonian is
\begin{equation}
{\cal H}=-\sum_{ij}t_{ij}c_{i}^{\dagger}c_{j}
+U\sum_{i}c_{i}^{\dagger}c_{i}f_{i}^{\dagger}f_{i}
-\mu\sum_{i}c_{i}^{\dagger}c_{i}+E_f\sum_if_{i}^{\dagger}f_{i},
\label{H}
\end{equation}
where $t_{ij}$ is the nearest-neighbor hopping matrix,
$U$ is the on-site repulsion between
$c$ and $f$ electrons, $\mu$ is the chemical potential of the conduction
electrons and $E_f$ is the site energy for the localized electrons.
In the simplest case, we ignore the spin of the electrons and assume they
are spinless. In the calculations presented here, we set $\mu=U/2$,
$E_f=-U/2$, so that $\langle c^\dagger c\rangle=\langle f^\dagger f\rangle
=1/2$; this case is called half filling.

The electric field ${\bf E}({\bf r}, t)$ is described by a vector potential
${\bf A}({\bf r}, t)$ in the Landau gauge where the scalar potential vanishes:
\begin{equation}
{\bf E}({\bf r}, t)=-\frac{1}{c}\frac{\partial {\bf A}({\bf r}, t)}
{\partial t}\, .
\label{eq: electric field}
\end{equation}
We assume that the vector potential ${\bf A}({\bf r}, t)$
is smooth enough in space, that the
magnetic field produced by ${\bf A}({\bf r}, t)$
can be neglected.

The electric field is introduced into the Hamiltonian (\ref{H})
by the so-called Peierls' substitution \cite{peierls,Jauho} where we neglect
interband transitions because we are considering only a single band (the 
possible dipole transition from a localized electron state to a
conduction electron state is also neglected; this assumption may break
down if the localized particles are electrons, but it cannot break down if 
they are ions as in a binary alloy interpretation of the FK model):
\begin{equation}
t_{ij}\rightarrow t_{ij}\exp\left[
-\frac{ie}{\hbar c}\int_{{\bf R}_{i}}^{{\bf R}_{j}}{\bf A}({\bf r}, t)\cdot
d{\bf r} \right] .
\end{equation}
The Peierls' substitution represents the effect of the line integral
of the vector potential on the hopping between sites $i$ (at position
${\bf R}_i$) and $j$ (at position ${\bf R}_j$); in this work $t_{ij}\ne 0$
only for nearest-neighbor sites $i$ and $j$.

For simplicity we shall study the case of a $d$-dimensional 
hypercubic lattice in the limit of large spatial dimensions
$d\rightarrow \infty$. In this limit, the electron self-energy
becomes local, which simplifies both the formalism and the numerical
calculations. This approximation corresponds to the dynamical
mean-field theory (DMFT) limit \cite{metzner_vollhardt}.
The simplest electric field is one that
lies along the unit cell diagonal \cite{Turkowski}:
\begin{equation}
{\bf A}(t)=A(t)(1,1,...,1).
\end{equation}
After the Peierls' substitution, the ``band-structure'' in the electric field
becomes
\begin{equation}
\epsilon_{\bf k}=
-2t\sum_{l}\cos\left[a\left( {\bf k}_{l}-\frac{e{\bf A}_{l}(t)}{\hbar c}
\right)\right]\, ,
\end{equation}
with $a$ the lattice spacing which we will take to be one.
With our choice for the electric field along the diagonal, this becomes
\begin{equation}
\epsilon_{\bf k}=
\cos\left(\frac{eA(t)}{\hbar c} \right)\varepsilon_{\bf k}
+\sin\left(\frac{eA(t)}{\hbar c} \right)\bar\varepsilon_{\bf k},
\label{energy}
\end{equation}
with
\begin{equation}
\varepsilon_{\bf k}=-\frac{t^{*}}{\sqrt{d}}\sum_{l}\cos k_{l}
\label{eps}
\end{equation}
and
\begin{equation}
{\bar \varepsilon}_{\bf k}=-\frac{t^{*}}{\sqrt{d}}\sum_{l}\sin k_{l}.
\label{bareps}
\end{equation}
being generalized energy functions and $t^{*}$ is a renormalized
hopping parameter:
$t=t^{*}/2\sqrt{d}$ in the limit $d\rightarrow\infty$~\cite{metzner_vollhardt};
$t^*$ will be used as our energy unit.

We find that many quantities we want to determine involve a summation over
momenta of functions of $\varepsilon$ and $\bar\varepsilon$.  These summations
can be performed more easily by determining a joint density of states for the
two energies in Eqs.~(\ref{eps}) and (\ref{bareps}); the result in the
limit of the infinite dimensions~\cite{Schmidt} becomes:
\begin{eqnarray}
\rho_{2}(\varepsilon , {\bar \varepsilon})
=\frac{1}{\pi t^{*2}}\exp\left[
-\frac{\varepsilon^{2}}{t^{*2}}-\frac{{\bar \epsilon}^{2}}{t^{*2}}
\right] .
\nonumber
\end{eqnarray}
Hence, a summation over an infinite-dimensional Brillouin zone can be
re-expressed as a two-dimensional Gaussian integral.

\begin{figure}[h!]% fig 1
\centering
\epsfxsize=2.8in
\epsffile{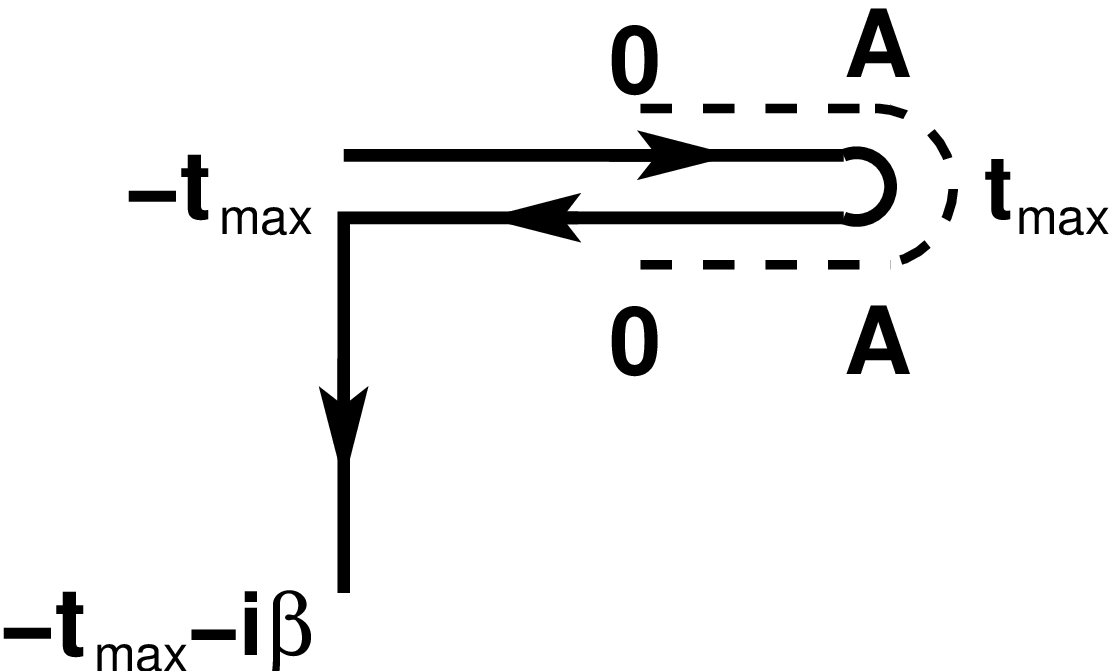}
\caption{Kadanoff-Baym 
integration contour for the time variables. The time-domain cutoffs
are symmetric at $\pm t_{\rm max}$. The direction for the integration of the
line integral is
indicated by the arrows.  The dashed line schematically shows where
we typically turn on the electric field, as represented by the
vector potential; it is commonly turned on when the time is equal to zero.
Note that for the lesser functions, we choose the first time argument on the
upper real time branch, and the second time argument on the lower real branch.
When the contour is discretized, we use a step spacing of $\Delta t$ along
the real axis, and a step size of 0.05 along the imaginary axis. All
calculations presented here have $\beta=1$ corresponding to twenty steps along
the imaginary axis.
\label{fig: contour}}
\end{figure}

In order to solve the many-body problem, we need to determine the electronic
Green's functions in the presence of the electric field.  The derivation
of formulas for
these Green's functions is more complicated than in the absence of a field,
because there is no time-translation invariance, so the Green's functions
depend on two different time arguments. Furthermore, since the system evolves
in the presence of an electric field, there is no simple way to relate
the quantum-mechanical state at large times to the state at small times.
Hence, we evolve the system forward in time, then we de-evolve it backwards
in time, in order to properly determine its complete
time evolution. Since the local
$f$-electron number is conserved, and we are not coupling the $f$ electrons
to the field, the Hamiltonian is a quadratic function of the conduction
electron operators.  This means the time-ordered product can be 
directly evaluated, and relevant functional derivatives can be taken
to determine the Green's functions.  The algebra is somewhat long and
will be omitted here.  The end result is a series of equations for the
so-called local contour-ordered Green's function, which is defined with 
two time arguments, each one lying on the Kadanoff-Baym contour
(see Fig.~\ref{fig: contour}):
$g^{c}(t,t')=-(i/\hbar)\langle \mathcal{T}_{c}c_i(t)c_i^{\dagger}(t')\rangle$,
with the time-ordering taking place along the contour, the time dependence of 
the fermionic operators being determined by the total (time-dependent)
Hamiltonian in the Heisenberg picture, and the angular brackets
denoting a weighted trace over all states $\langle \mathcal{O}\rangle={\rm Tr}
e^{-\beta\mathcal{H}}\mathcal{O}/\mathcal{Z}$; the partition function is
$\mathcal{Z}={\rm Tr} e^{-\beta\mathcal{H}}$ with $\beta=1/T$ the
inverse temperature. In addition, we need to define a
local self-energy $\Sigma^{c}(t,t')$ and an effective dynamical
mean-field $\lambda^{c}(t,t')$ in analogy with
the equilibrium case in zero electric field~\cite{brandt_mielsch,Freericks}:
\begin{equation}
g^{c}(t,t')=\int d\varepsilon \int d{\bar \varepsilon}
\rho_{2}(\varepsilon ,{\bar \varepsilon})
\left[g_{0}^{c-1}(\varepsilon,{\bar \varepsilon})
-\Sigma^{c}
\right]^{-1}(t,t')
\label{GF2set}
\end{equation}
\begin{equation}
\lambda^{c}(t,t') = g_{imp}^{c-1}(t,t')-g^{c-1}(t,t')-\Sigma^{c}(t,t')
\label{lambdaset}
\end{equation}
\begin{eqnarray}
g^{c}(t,t')&=&[1-w_{1}][g_{imp}^{c-1}[\mu]-\lambda^{c}]^{-1}(t,t')\nonumber\\
&+&w_{1}[g_{imp}^{c-1}[\mu\rightarrow\mu -U]-\lambda^{c}]^{-1}(t,t')
\label{gc3set}
\end{eqnarray}
\begin{equation}
\Sigma^{c}(t,t')=g_{imp}^{c-1}(t,t')-g^{c-1}(t,t')-\lambda^{c}(t,t'),
\label{Sigmaset}
\end{equation}
where $g_{0}^{c}(\varepsilon,{\bar \varepsilon},t,t')$
is the noninteracting Green's function
in the presence of the electric field~\cite{Turkowski},
$g_{imp}^{c}(t,t')$ is the impurity Green's function in zero field,
which is equal to $g_{0}^{c}(\varepsilon,{\bar \varepsilon},t,t')$
at $\varepsilon={\bar \varepsilon}=0$, and $w_{1}$ is the occupancy 
of the $f$-electrons [$w_1=\langle f^\dagger f\rangle]$.
The difference of this system of equations
from the real-frequency case is that these objects are all
continuous square matrix operators of time
(defined on the Kadanoff-Baym time contour) rather than being scalar
functions of frequency. The inverses are all to be interpreted as
matrix inverses.

We will be interested in the so-called lesser Green's function $g^<$
and self-energy $\Sigma^<$
in this work.  These functions are extracted from the contour-ordered
objects by fixing the first time argument $t$ to lie on the upper real-time
piece of the Kadanoff-Baym contour and the second time argument $t'$ to
lie on the lower real-time piece of the Kadanoff-Baym contour.

The system of the equations (\ref{GF2set})--(\ref{Sigmaset})
can be solved by iteration starting
from some initial guess for the self-energy $\Sigma^{c}(t,t')$.
From Eq.~(\ref{GF2set}), one can find the local Green's function
$g^{c-1}(t,t')$, which allows us to find the effective dynamical
mean field $\lambda^{c}(t,t')$ from Eq.~(\ref{lambdaset}).
Then the impurity equation [Eq.~(\ref{gc3set})] allows us to find a new
local Green's function, which is employed to find a new self-energy
from Eq.~(\ref{Sigmaset}). This procedure is repeated until
the self-energy has converged to a fixed point. We call this iterative solution
approach the DMFT algorithm.

Once the system of equations in Eqs.~(\ref{GF2set})--(\ref{Sigmaset}) is solved,
then we can determine the properties of the system as a function of time.
It is convenient to describe response functions like the self-energy
and the Green's function with
the relative $t_{\rm rel}$ and the average $T$
time variables (the so-called Wigner coordinates~\cite{wigner}):
\begin{equation}
t_{\rm rel}=t-t', \ \ \ \ T=\frac{t+t'}{2},
\end{equation}
instead of the two times $t$ and $t'$.
In equilibrium, these functions are independent of the average time $T$,
and depend only on the relative time $t_{\rm rel}$.  We perform a Fourier
transform of the relative time to a real frequency, and examine the 
Green's function and self-energy as functions of frequency.  In the
nonequilibrium case, we do a similar thing, performing the Fourier transform
with respect to the relative time, and examining how the response functions
evolve as a function of the average time. For example, we have
\begin{equation}
\Sigma^{<}(\omega ,T)=\int dt_{\rm rel} e^{i\omega t_{\rm rel}} 
\Sigma^{c}\left ( T+\frac{t_{\rm rel}}{2},T-\frac{t_{\rm rel}}{2}\right ).
\end{equation}
Note that we must have the first time argument $t=T+t_{\rm rel}/2$ lying on the
upper branch and the second time argument $t'=T-t_{\rm rel}/2$ lying on the
lower branch of the Kadanoff-Baym contour.

Another interesting quantity is the current density that is driven by the
external electric field:
\begin{equation}
{\bf j}_l(T)=
-i\frac{et^{*}}{\sqrt{d}}
\sum_{{\bf k}}\sin \left(
{\bf k}_{l}-\frac{e{\bf A}_{l}(T)}{\hbar c}
\right)
g^{<}(\varepsilon_{{\bf k}},\bar\varepsilon_{\bf k},T,T),
\end{equation}
with each vector component identical when the electric field lies along the
diagonal.  The 
magnitude of the total current density in the case where the electric field
lies along the unit-cell diagonal is then
\begin{equation}
j(T)=\sqrt{d}{\bf j}_{l}(T).
\end{equation}

It is well known that the current response to an electric field is strange
for a perfect conductor that has no electron scattering. Indeed, there
is an $ac$ response to a $dc$ field due to the lattice periodicity, which
does not allow the momentum of the electron to get too large before an
umklapp scattering event occurs with the lattice and changes the sign
of the momentum.  This phenomenon is called a Bloch 
oscillation~\cite{bloch,zener,a_m}, and
it should be seen in any material that is free enough of defects
and other sources of scattering.  No
conventional metal has ever been grown that has small enough scattering to
exhibit Bloch oscillations.  Instead, the scattering occurs so rapidly, that
the steady-state current is a constant, which increases linearly with the 
electric field until nonlinear effects take over. Bloch oscillations have been
seen in semiconducting heterostructures~\cite{bloch_semi}

Bloch oscillations are also seen in DMFT, with a
time-independent electric field ($E$ constant and $A(T)=-EcT$)~\cite{Turkowski}:
\begin{equation}
j(T)\sim
\sin \left(\frac{eA(T)}{\hbar c}
\right)
\int d\varepsilon
\frac{df(\varepsilon -\mu )}{d\varepsilon}
\rho (\varepsilon) ,
\end{equation}
producing an oscillating current density [$\rho(\varepsilon)$ is the 
noninteracting density of states, which is equal to the integral of
$\rho_2$ over $\bar\varepsilon$ and $f(\varepsilon)=1/\{1+\exp(\beta\varepsilon)
\}$ is the Fermi-Dirac distribution].
The frequency of the oscillation is $\omega_{\rm Bloch}=eE/\hbar$ and
is called the Bloch oscillation frequency. We expect these
oscillations to survive in the presence of scattering
if the field is large enough that the relaxation time due to scattering
is significantly larger than the Bloch oscillation period. The frequency
of oscillation is undoubtedly too high for the Bloch oscillations
to be directly observed  ($\omega_{\rm Bloch} \gg 10^{12}$~Hz).

\section{Numerical algorithm}

There are a significant number of numerical issues that need to be taken
into account to be able to determine the Green's functions, and other
properties of strongly correlated electrons in a large electric field.
To begin, the matrix operators are continuous operators defined along the
Kadanoff-Baym contour, and there is no simple way to find their matrix
inverse analytically.  Furthermore, matrix multiplication implies an
integration over the Kadanoff-Baym contour
\begin{equation}
A\cdot B(t,t')=\int_c dt'' A(t,t'')B(t'',t'),
\end{equation}
which is a complicated line integral in the complex plane (see 
Fig.~\ref{fig: contour}).  Our approach
to solve this problem is a common numerical approach---we discretize the
Kadanoff-Baym contour and evaluate the line integrals as finite Riemann sums 
over the discretized paths.  The matrix operators then become finite-dimensional
square matrices, whose size is equal to the number of points used to
discretize the Kadanoff-Baym contour. Once this has been accomplished, then
standard LAPACK and BLAS routines can be employed to invert and manipulate
the discretized versions of the matrix operators. One issue that needs to
be taken into account though is that the inverse of a continuous matrix
operator satisfies
\begin{equation}
\int_c dt'' A^{-1}(t,t'')A(t'',t)=\delta_c(t,t')
\end{equation}
with $\delta_c$ the Dirac delta function on the contour. The delta function
is represented by the inverse of the time step used in the discretization
of the Kadanoff-Baym contour, but one needs to note that this time step changes
sign on the lower (real) branch of the contour, and it becomes imaginary
on the vertical piece of the contour. One needs to properly take this into
account before using a matrix inversion routine.

Next, we need to examine the numerical issues arising in Eq.~(\ref{GF2set}).
We evaluate this equation for a given self-energy matrix $\Sigma^c(t,t')$.
This requires us to choose values of $\varepsilon$ and $\bar\varepsilon$
for the two-dimensional Gaussian integration, compute the inverse of the
matrix $g_0^c(\varepsilon,\bar\varepsilon)$, subtract the self-energy,
and compute a new matrix inverse. There is an exact algorithm that allows us to
directly compute the matrix inverse of $g_0^c(\varepsilon,\bar\varepsilon)$;
this arises from the equation of motion for the Green's function, from which
the inverse operator can be directly read off.  The only subtlety is to ensure
that the inverse operator inherits the correct boundary condition from
the Green's function.  This is not so simple to carry out, but using techniques like the discretization scheme in Negele and Orland~\cite{negele_orland}
and Kamenev~\cite{kamenev}
provides a systematic method to directly compute the matrix inverse which
satisfies the requisite boundary condition and becomes the exact matrix-operator
inverse in the limit where the discretization step size goes to zero.
The last matrix inversion is a general complex
matrix inversion, because the self-energy is complex-valued, and has no
simple symmetries.  Hence, it is the most ``expensive'' 
matrix inversion that needs
to be performed.  Next the matrix elements of the inverse are multiplied by the
relevant weighting factors for the integration, and finally we accumulate the 
results over all $\varepsilon$ and $\bar\varepsilon$ terms that we choose for
the two-dimensional integral.  Since the integral weights are Gaussian, it
seems reasonable to employ a Gaussian integral scheme for choosing the points
on the grids and the weights.  Unfortunately, since each point in the
two-dimensional integration requires one full matrix inversion, we need to
minimize the number of points chosen.  As a compromise, we use the following 
scheme: (i) we perform the integral using an $N=54$ Gaussian integration;
(ii) repeat with an $N=55$ Gaussian integration; and (iii) average the
results. The $N=54$ case requires $2916$ grid points and the $N=55$ case 
requires
$3025$ grid points. We choose to average these two results, because terms
in the Green's function often behave like $\exp(ic\varepsilon)$, which can be
accurately represented by the Gaussian integration until $c$ becomes on the
order of the inverse of the grid spacing of the Gaussian integration near
$\varepsilon=0$. Then, the sampling over the discrete points will no longer
cancel, and the Gaussian integration will overestimate the value of the
integral.  Since the grid spacing for $N+1$ points nearly interlaces that for 
$N$, the results of the averaging over $N$ and $N+1$ grids produces accurate
results for values of $c$ up to two times larger than what is possible
for either one alone, and in the double integral case, it produces a 
factor of two reduction in computation time from using a Gaussian integration
scheme with twice the number of points.

This part of the DMFT algorithm, the calculation of the local Green's function 
from the self-energy, is easily parallelized.  One simply ships the 
self-energy matrix, and the energy variables $\varepsilon$ and
$\bar\varepsilon$ to the individual nodes, generates the relevant matrix,
performs the inversion, and sends the result back 
to the master node for accumulation.  Once the local Green's
function has been calculated, then we proceed with the remainder of the DMFT
algorithm to determine the new self-energy matrix.  This part of the code
is not so simple to parallelize, because it must proceed in a serial fashion.
The only possible parallelization will occur if we can use SLAPACK routines
to distribute the calculation of the matrix inverses over a small set of 
processors. To date, we have not included this element in the computation.

The algorithm given by Eqs.~(\ref{GF2set})--(\ref{Sigmaset}) is iterated
until it converges.  As the size of the matrices is made larger, by choosing
a larger maximal cutoff in time for the Kadanoff-Baym contour, or by
fixing the maximal cutoff time and reducing the step size, then the algorithm
slows down significantly, and it becomes more difficult to attain the same level
of accuracy at the end of the iterations.  We usually try to iterate the 
solutions at least $20-50$ times for full convergence, but sometimes we have to
limit ourselves to about $10$ iterations due to the computational time
involved.

\begin{figure}[t!]% fig 2
\centering
\epsfxsize=2.83in
\epsfclipon
\epsffile{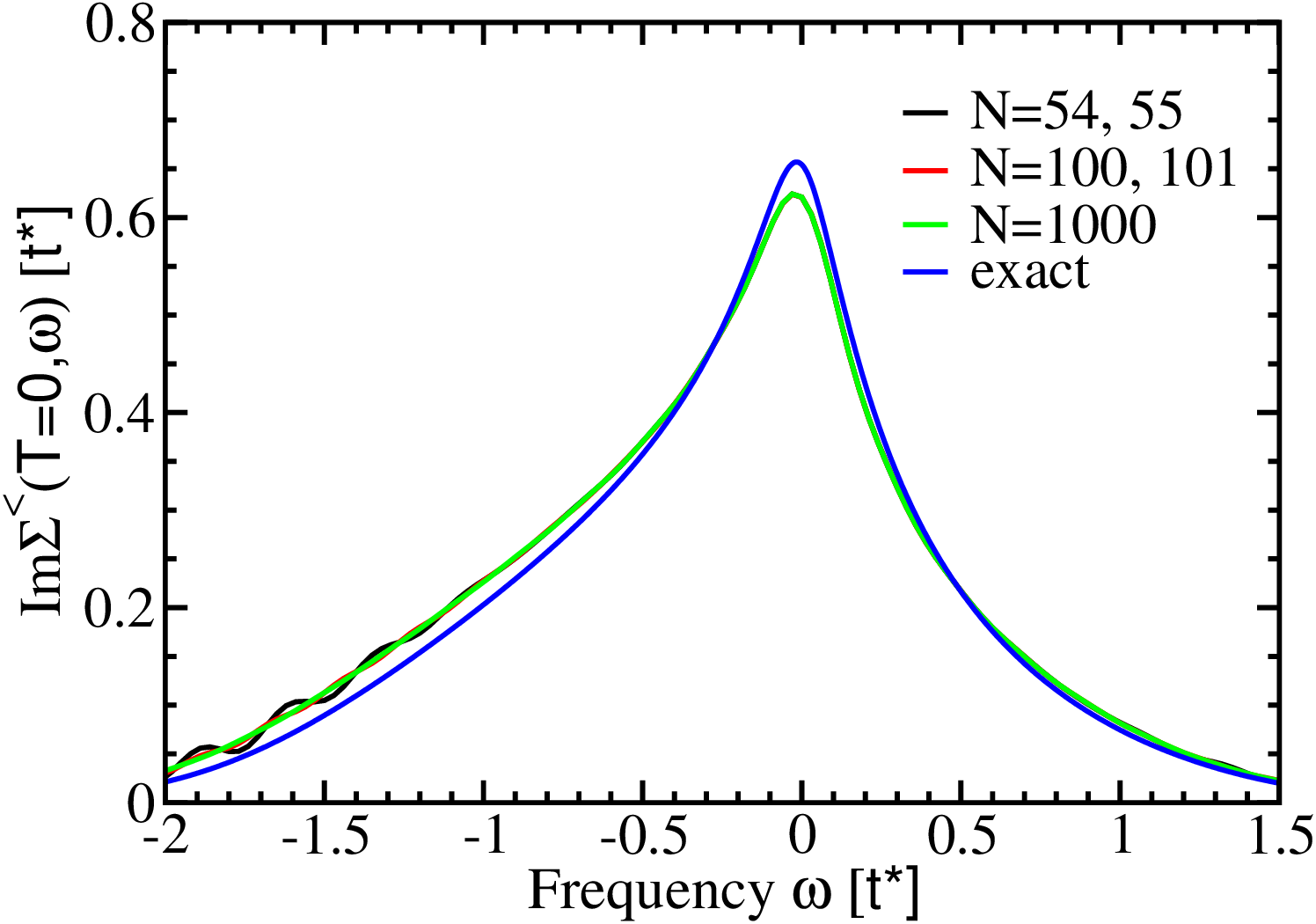}
\caption{Lesser self-energy for $\beta=1$, $U=1$, and half-filling.
We examine the frequency dependence of the self-energy at an average
time $T=0$ and in equilibrium, but calculated with the nonequilibrium
formalism.  The exact results are in blue, and the other curves are all
calculated with $\Delta t=0.05$ and $t_{\rm max}=15$.  The black curve uses
Gaussian integration with $N=54$ and $55$ points.  The red curve is for
$N=100$ and $101$ points, and the green curve is a trapezoidal rule with
$1000$ points evenly spaced between $-3$ and $3$.
\label{fig: sigma_w}}
\end{figure}

\begin{figure}[h!]% fig 3
\centering
\epsfxsize=2.83in
\epsfclipon
\epsffile{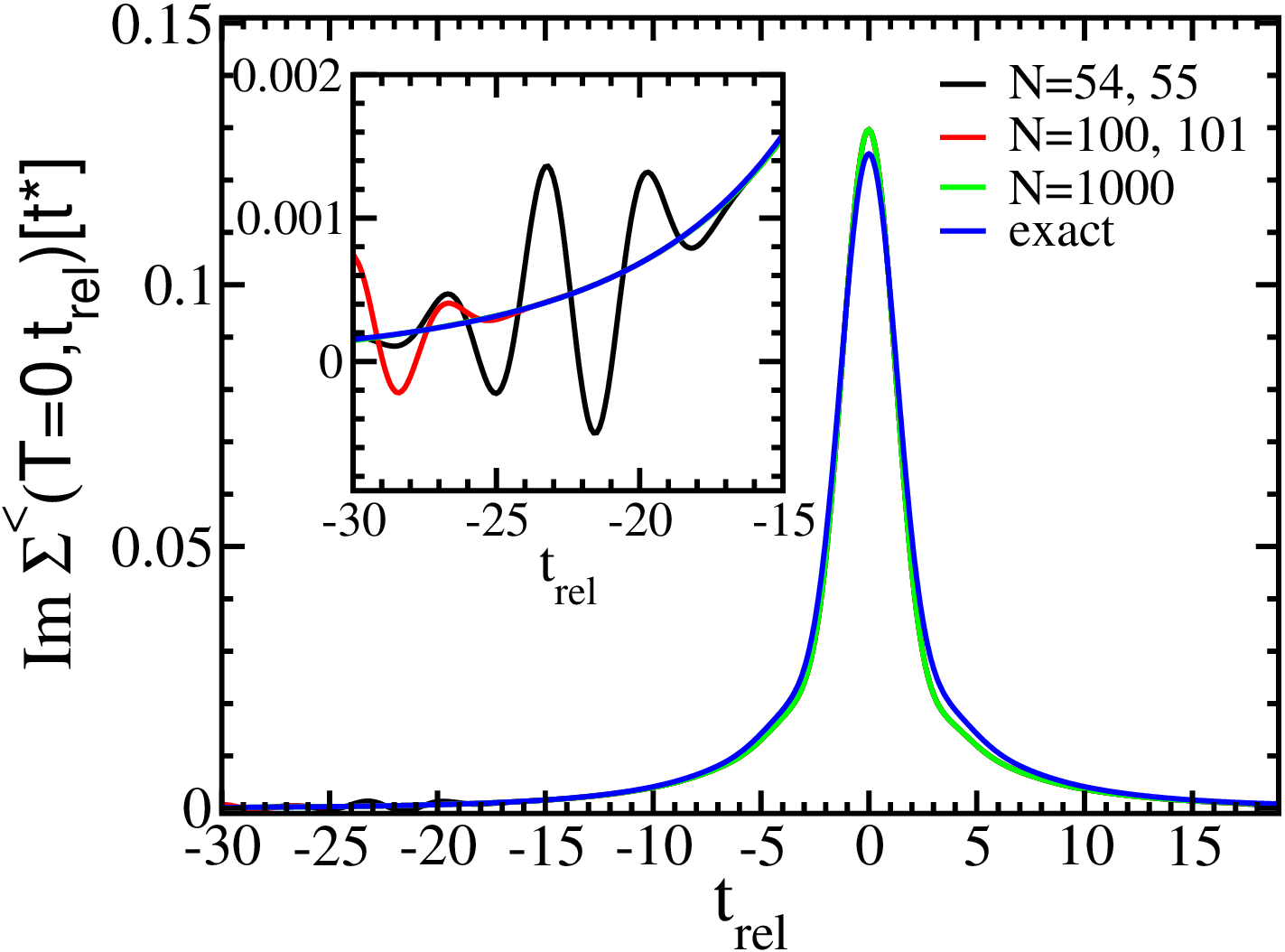}
\caption{Imaginary part of the 
lesser self-energy for $\beta=1$, $U=1$, and half-filling.
We examine the relative time dependence of the self-energy at an average
time $T=0$ and in equilibrium, but calculated with the nonequilibrium
formalism.  The exact results are in blue, and the other curves are all
calculated with $\Delta t=0.05$ and $t_{\rm max}=15$. The colors of the 
other curves are identical to those in Fig. 2. Inset is a blow up of the
region around $t_{\rm rel}=-20$, where the integration over $\varepsilon$
introduces spurious results when the grid spacing is too coarse, which
lead to the oscillations seen in Fig.~2.
\label{fig: sigma_t}}
\end{figure}

\begin{figure}[h!]% fig 4
\centering
\epsfxsize=2.83in
\epsfclipon
\epsffile{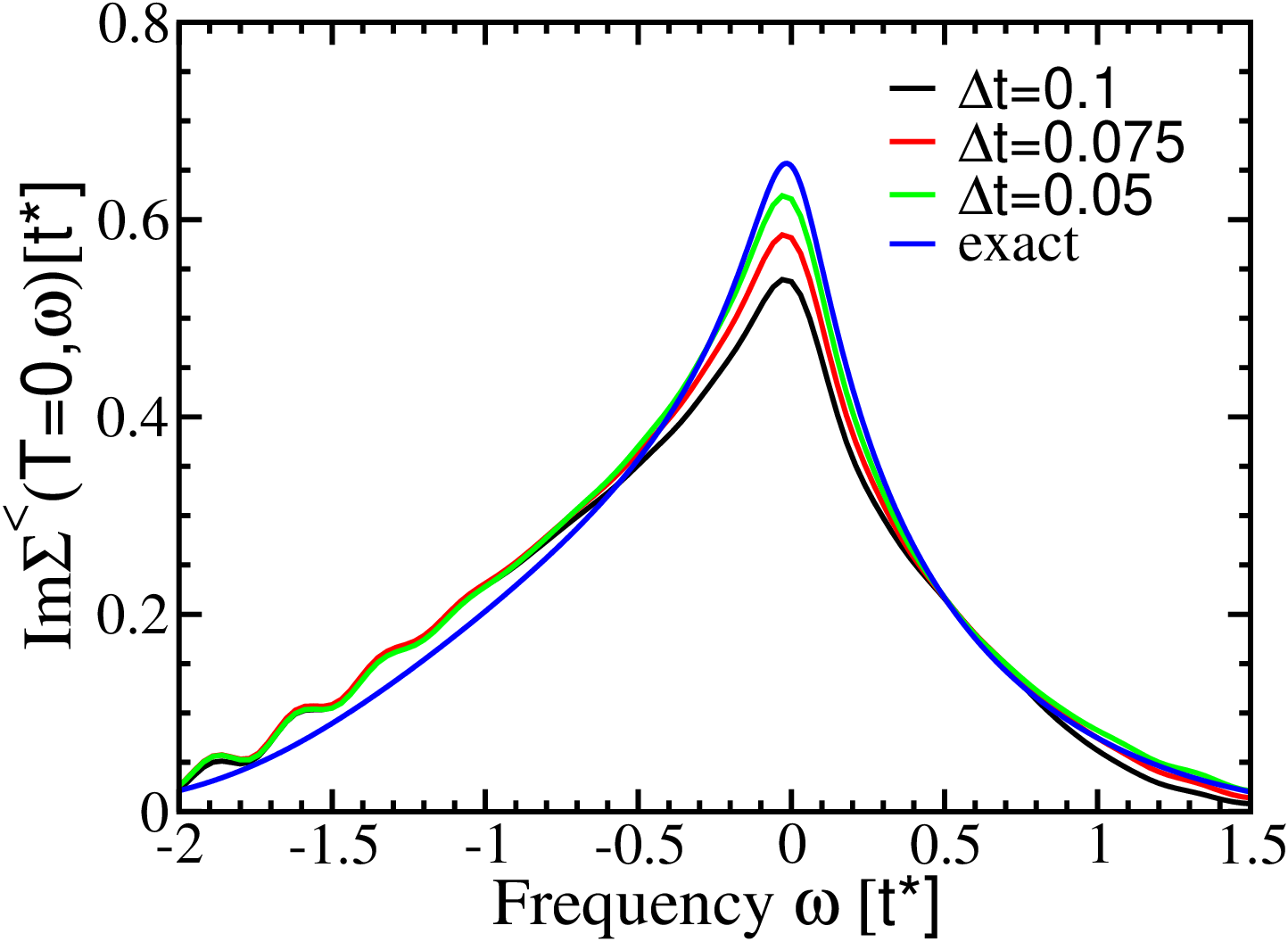}
\caption{Lesser self-energy for $\beta=1$, $U=1$, and half-filling for
different discretizations of the Kadanoff-Baym contour.
The exact results are in blue, and the other curves are all
calculated with $N=54$  and $55$ points for the Gaussian integration
and $t_{\rm max}=15$.  The black curve 
has $\Delta t=0.1$.  The red curve has $\Delta t=0.075$
and the green curve has $\Delta t=0.05$.
\label{fig: sigma_w2}}
\end{figure}

\begin{figure}[h!]% fig 5
\centering
\epsfxsize=2.90in
\epsfclipon
\epsffile{fig5.eps}
\caption{Lesser Green's function for $\beta=1$, $U=1$, and half-filling for
different discretizations of the Kadanoff-Baym contour. We plot
the results for $T=0$ as a function of $t_{\rm rel}$. The main panel is
the imaginary part, and the inset is the real part.
The exact results are in blue, and the other curves are all
calculated with $N=54$  and $55$ points for the Gaussian integration
and $t_{\rm max}=15$.  The black curve
has $\Delta t=0.1$.  The red curve has $\Delta t=0.075$
and the green curve has $\Delta t=0.05$. Note that the exact result is
an even function of $t_{\rm rel}$ for the imaginary part and an odd
function for the real part.
\label{fig: gf_t}}
\end{figure}

\begin{figure}[h!]% fig 6
\centering
\epsfxsize=2.90in
\epsfclipon
\epsffile{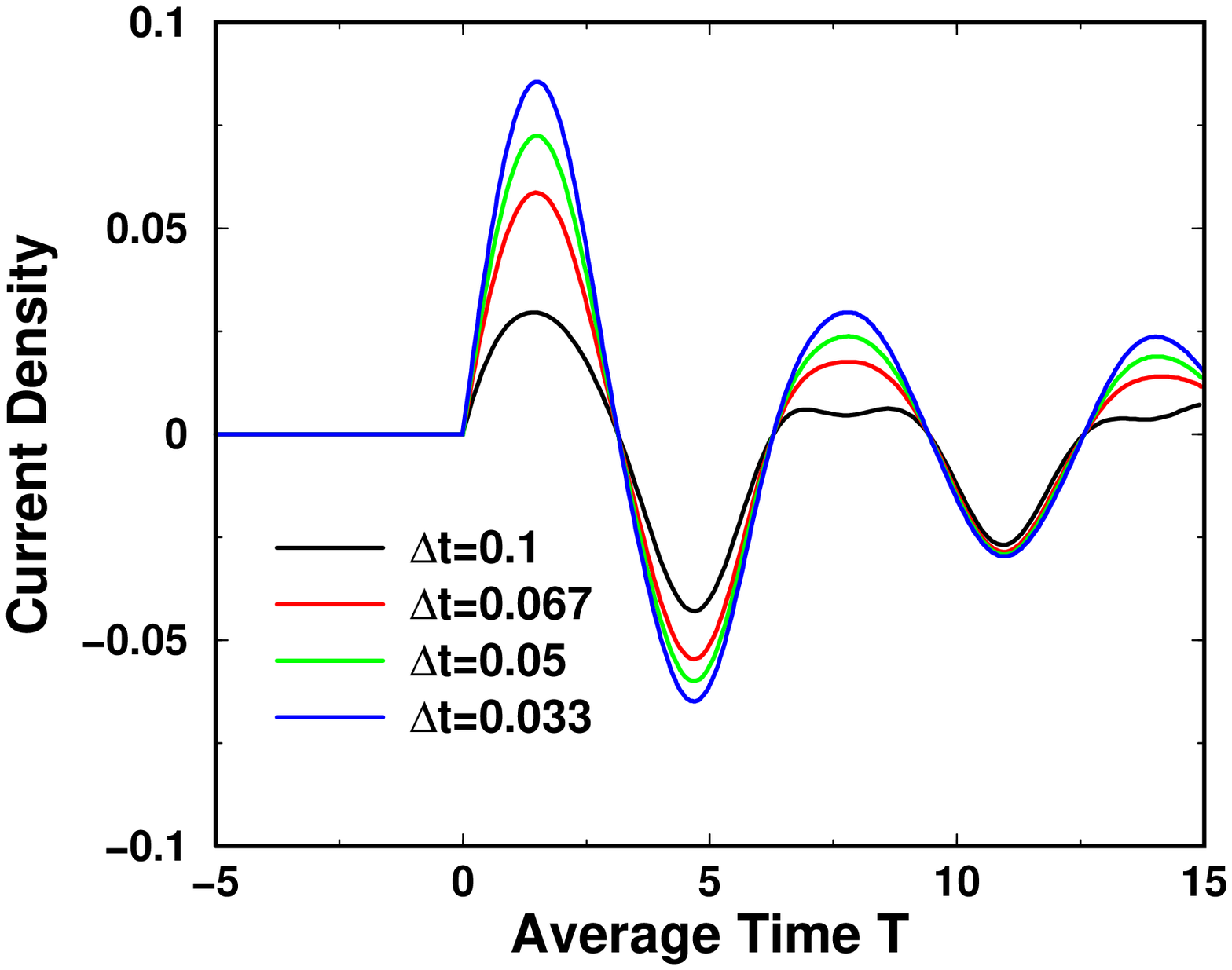}
\caption{Current density as a function of time for $U=0.5$ and various
values of $\Delta t$ ($t_{\rm max}=15$ and $N=54$, $55$ Gaussian
integration).  The temperature is fixed at $\beta=1$ and the
electric field satisfies $E=1$.  Note how the current converges to a 
``fixed point'' as $\Delta t\rightarrow 0$, but not uniformly.
\label{fig: current_0.5}}
\end{figure}

\begin{figure}[h!]% fig 7
\centering
\epsfxsize=2.90in
\epsfclipon
\epsffile{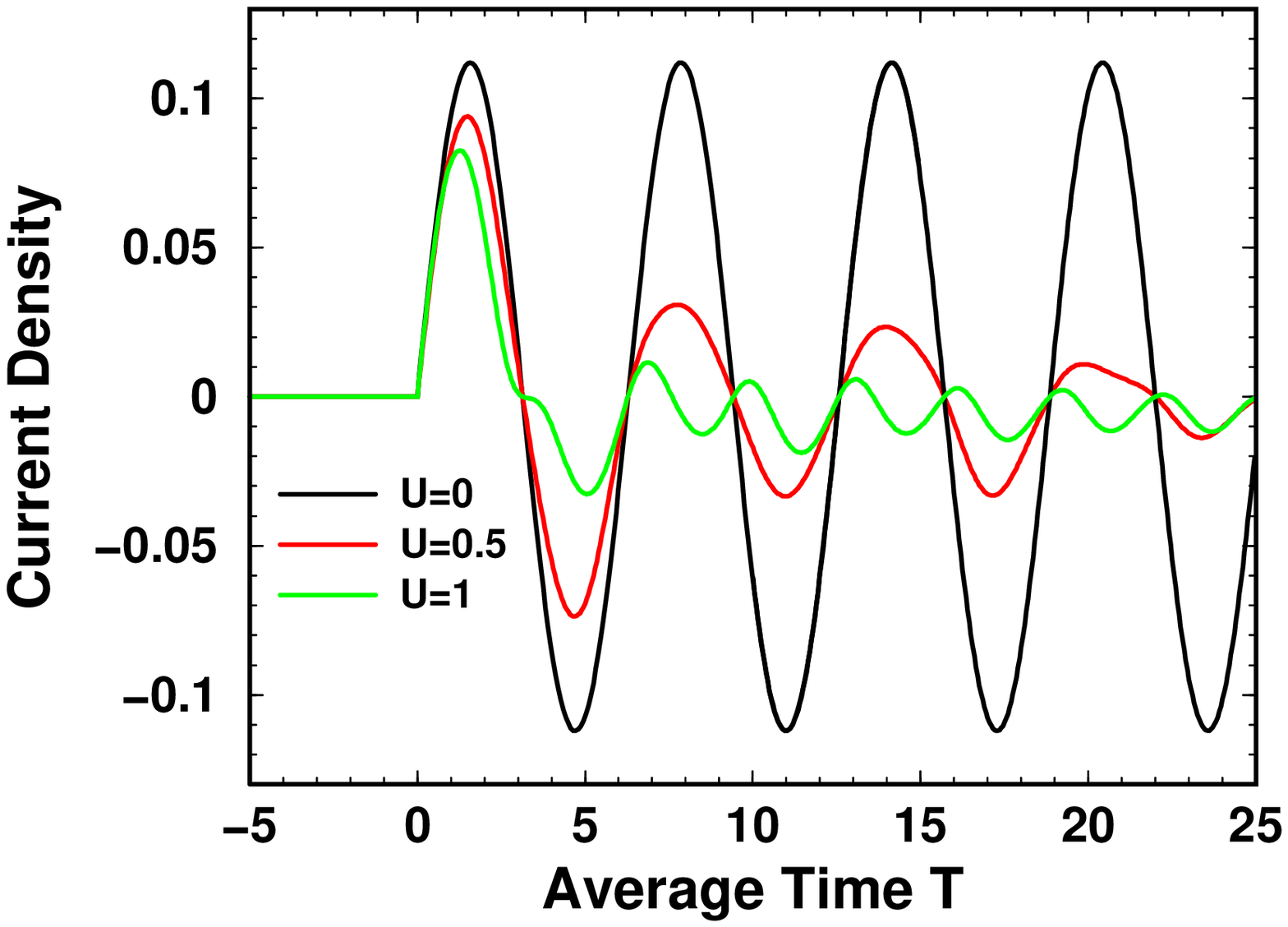}
\caption{Current density as a function of time for $U=0$, $0.5$, and $1$.
The temperature is fixed at $\beta=1$ and the
electric field satisfies $E=1$ ($N=54$, $55$ Gaussian integration).  
Note how the Bloch oscillations are
damped as the scattering increases, and how the period might be modified.
\label{fig: current_u}}
\end{figure}

There are a number of numerical issues that play a role in the quantitative
accuracy of the results.  These include the maximal time chosen for the
cutoff $t_{\rm max}$, the step size in real time $\Delta t$, and the 
number of points $N$ chosen for the $\varepsilon$ grids. In this work, we
examine a number of different choices for these parameters to see which
terms are the most important for maintaining accuracy of the results.
We do this employing what is probably one of the most difficult cases
for the nonequilibrium code, that of a vanishing electric field (corresponding
to an equilibrium situation).  There is a major simplification of this approach
because we have no $\bar\varepsilon$ dependence in our formulas, and the
integral over $\bar\varepsilon$ can be done trivially.  The problem arises
from the fact that the self-energy develops a delta function in frequency
at $\omega=0$ in the insulating state, and this function cannot be easily
represented in a calculation for real time that has a finite cutoff along
the time axis.  Indeed, we find that it is difficult to obtain good results
for the self-energy as a function of frequency in this case. But we also
check the moments of the self-energy and the Green's function, and find
good agreement for the low-power moments, indicating that the results
do work well for short times.  Hence, it is reasonable to think that they
do a good job at determining the initial transient response to an electric field
after the field is turned on.  Furthermore, we do not know whether the
delta function in the self-energy survives when a field is turned on.  If
it doesn't, then the nonequilibrium approach may work well for the insulating
cases, even if it has problems for the equilibrium system.

\section{Benchmarking the equilibrium results}

The electronic density of states is a quantum-mechanical measure of the
elementary excitations of a strongly correlated material.  In most materials,
the density of states spans a range of a few electron volts, or in our
units, a few $t^*$. The Fourier transform of the density of states is
closely related to the Green's functions in the time domain.  We expect 
the time-dependent Green's functions to have important time dependence when 
$t_{\rm rel}$ is smaller than $c/t^*$ for some constant $c$ of order one.
Hence choosing a cutoff in time on the order of $10-20$ $1/t^*$ is a reasonable
choice.  The above heuristic discussion holds when the system is metallic.
When it becomes insulating, then the density of states has a gap, and the 
self-energy develops a pole, for small frequencies.  In this case, we need a 
large time domain (infinite in the case of a delta function) to properly
reproduce the density of states or self-energy
after Fourier transforming.  This means that
the time-domain cutoff needs to be made larger as the electron correlations
are made stronger.

Similarly, the more fine-structure that is present in the Green's functions
as functions of time, the smaller the step size $\Delta t$ in the time domain
needs to be to be able to accurately discretize the matrix operators.  In
general, we expect the discretization error to grow as the electron correlations
increase, so $\Delta t$ will need to be reduced for larger values of $U$.
We typically choose $\Delta t=0.1-0.05$ in our calculations.  Obviously,
we cannot keep increasing the time domain, and decreasing the step size
when we have finite computational resources.  The maximal matrix that we 
consider has a size on the order of $2500\times 2500$.  This choice is
not a function of memory limitations, but rather is an issue of the
computational time needed to invert these matrices as part of the DMFT
algorithm. The bottom line is that these calculations will not be able
to be pushed to too large a value for $U$ without running into resource
problems.

In order to benchmark our code, we have chosen to examine the equilibrium
solutions using the nonequilibrium formalism.  This is a particularly
nice exercise to undertake, because the equilibrium solutions are all
known to high accuracy via a direct solution using an algorithm in
the frequency domain~\cite{brandt_mielsch,Freericks}. 
It is also a challenging test of the nonequilibrium codes,
because we need to Fourier transform the solutions in time to functions
of frequency, and the effects of the discretization step size, and of the 
time-domain cutoff can play critical roles in the accuracy.  The other
important parameter in governing the accuracy is the step size employed
in the energy integrations to determine the local Green's function. 
As discussed above, if the step size is too large, then we can generate 
spurious signal at large relative times, which will Fourier transform
into high-frequency wiggles in the frequency domain.  If we know that
such structures are not present in the exact results, we can easily
filter them out, but this becomes problematic when we are not sure
whether such structures are real or numerical artifacts (which occurs when
we perform nonequilibrium calculations).

We benchmark our results by examining the equilibrium solutions at high
temperature ($\beta=1$) and for a large value of $U$ ($U=1$) that is still in 
the metallic phase (the metal-insulator transition occurs at $U=\sqrt{2}$).
We choose this for our initial benchmarking exercise because the self-energy
does not have a pole; hence, the numerical issues should be under better
control.  We will briefly discuss issues that occur in the insulating phase
below.

Our first result is shown in Fig.~\ref{fig: sigma_w}.  It plots the 
lesser self-energy at $T=0$ as a function of $\omega$, which is calculated
by performing the Fourier transformation with respect to $t_{\rm rel}$. This
result should be independent of $T$, because we are in equilibrium, but the
results do have a small dependence on $T$, that is due to the fact that we 
have instituted a finite cutoff in time $t_{\rm max}=15$. In this figure,
we study how sensitive the results are to changing the number of points
in the integration over $\varepsilon$ (recall, the integration over
$\bar\varepsilon$ is trivial when we are in equilibrium). The blue curve
is the exact result, the black curve employs Gaussian integration, averaging
over the $N=54$ and $N=55$ cases, while the red curve is similar with
$N=100$ and $N=101$ points. The green curve uses a much smaller step
size in $\varepsilon$, employing $N=1000$ points in a trapezoidal rule
integration, ranging from $-3$ to $3$.  Note how all of the results lie on
top of each other for small $\omega$, although they do differ from the exact
result.  In the tails, for larger $|\omega|$, we see oscillations
develop for the Gaussian integrations, that are reduced as the step size is made
smaller, and completely disappear by the time $N=1000$.  These results show that
by carefully controlling the step size used for the energy integration,
one can obtain converged results without any extraneous oscillations, but
those converged results are not exact, because they were calculated with 
finite values for $\Delta t$ and $t_{\rm max}$.

In Fig.~\ref{fig: sigma_t}, we show the results for the imaginary part of the
lesser self-energy as a function of $t_{\rm rel}$. Once can see particularly
good quantitative agreement for small times, and then there is a region
with oscillations out in the tail ($t_{\rm rel}\approx -20$).  
This region is blown up in the inset.
The oscillations are only present for the coarse $\varepsilon$ integrations,
and the amplitude of the oscillations shrinks as the step size is made smaller.
It is precisely these spurious results that lead to the oscillations in
the Fourier transform (see Fig.~\ref{fig: sigma_w}).  
If we know about this kind of spurious behavior,
we can filter the oscillations out before performing the Fourier transform,
but this is an {\it ad hoc} procedure that cannot be generalized to the
nonequilibrium case.

In Fig.~\ref{fig: sigma_w2}, we show the $\Delta t$ dependence of the 
calculations with fixed values for $t_{\rm max}=15$, and $N=54$ and $N=55$
averaged Gaussian integrations.  The results vary the most at small
frequency, and appear to systematically approach the exact result as $\Delta t
\rightarrow 0$.  The results are less sensitive to $\Delta t$ for larger 
frequencies, and since we have already seen that reducing the step size
in $\varepsilon$ tends to only smooth out the oscillations (without changing
the shape too much), the errors
at higher frequencies must be coming from the finite cutoff $t_{\rm max}$.

Finally, we show the Green's function as a function of $t_{\rm rel}$ in
Fig.~\ref{fig: gf_t}. The imaginary part is in the main plot, and the
real part is in the inset.  In the exact result (blue curve), the imaginary
part is an even function, and the real part is an odd function.  The
results of the nonequilibrium calculation do not share this symmetry, but it 
appears to be getting restored as $\Delta t\rightarrow 0$.  More problematic
is the issue of the value of $g^<$ at $t_{\rm rel}=0$, which is determined
solely by the electron filling.  The results appear to be getting worse
as $\Delta t\rightarrow 0$.  It must be that if we increase the
time cutoff, the results will ultimately start moving back toward their
correct value, but we cannot check this explicitly due to the finite
computer resources that are available.

We can be more quantitative about the short time behavior though.  To
do so, we can calculate the first few moments of the integrals of the
Green's function and the self-energy over $\omega$, and compare them to the
exact results for those integrals.  The zeroth moment relates to the function
at $t_{\rm rel}=0$, the first power to the slope, and the second power to the
curvature.  When we examine the results for $U=0.5$ and $U=1$, we find the 
zeroth moment of $g$ and $\Sigma$
is in error by about $7$\%, the first moment by $10$\% for $g$ and
$20$\% for $\Sigma$, and the second moment of $g$ and $\Sigma$ by $15-20$\%. 
The results do not depend
too strongly on the step size for $\varepsilon$ in the integrations, as 
expected, because the oscillations average out of the moments.  The first
moment appears to extrapolate to its correct value as $\Delta t\rightarrow 0$
for the $N=54$, $55$ Gaussian integration with a fixed $t_{\rm max}$, but
the zeroth and second moments do not appear as if they scale to the right 
result. This is most likely due to the fact that the $t_{\rm max}$ needs to
be increased, but the large step size for $\varepsilon$ may also play
a role.

When we try to examine the insulating phase, we find the agreement with
the exact results becomes much worse.  This is because there are
low-energy features which require large times to be determined accurately.
Also, the larger $U$, the smaller $\Delta t$ needs to be to obtain good
accuracy.  Our results for $U=1.5$ are too preliminary to report 
quantitative values here.

The important question is whether these low-energy features survive as
the electric field is turned on.  If they are destroyed by the field, then
the computational scheme that we are using should be able to accurately
determine nonequilibrium results at short times.  If they survive, then it
will be difficult to get high accuracy results for the nonequilibrium
case in the Mott insulator.  Since the presence of a field pumps energy
into the system, and that energy can be used to create excitations across
a gap, it is easy to believe that the gap features do not survive the
introduction of a large electric field, but we cannot definitively say
whether this is actually true at this point.
\section{Bloch oscillations}

One of the most interesting nonlinear phenomena of a material is the
production of Bloch oscillations in the current as a function of time
when a constant ($dc$) electric field is applied.  In the absence of
interactions (which cause scattering), the current will oscillate
forever, with a constant amplitude (the period is determined by the 
strength of the electric field). As we turn on the scattering, we expect the
oscillations to be dampened, but perhaps to maintain the same period (and
even survive in the steady state).  If
the scattering becomes large enough, then the oscillations should 
disappear completely.  By calculating the Green's functions in the
presence of a field that is turned on at $T=0$, we can study how the current
initially starts, and how it evolves into a steady state.  Due to the
need for a finite $t_{\rm max}$, we can only go so far out in time before
the calculation must terminate.

In Fig.~\ref{fig: current_0.5}, we plot the Bloch oscillations of the current
as a function of the average time, for $U=0.5$, $\beta=1$, $E=1$,
and the $N=54$, $55$
Gaussian integration scheme. The $t_{\rm max}$ is equal to 15. As $\Delta t$
is reduced, one can clearly see that the results are beginning to converge.
Furthermore, it is also clear that there is a damping of the transient
response as we move forward in time. We examine the behavior of this 
transient damping in Fig.~\ref{fig: current_u}, where we have preliminary
results for the current density for three values of $U$.  The results
for $U=0.5$ and $U=1$, are calculated with a step size of $\Delta t=0.05$.
At the moment, it is not clear how much of an effect the boundaries at 
$t_{\rm max}$ have on the results, but it may be that the largest time results
are not fully trustworthy.  Some interesting behavior can be seen in
the figure: (i) we see the damping that is expected, and that it gets
more strongly damped as $U$ is increased; (ii) as $U$ is increased, it appears
that the period may be decreasing, which is not an expected result; and (iii)
the behavior of the transient evolution is quite complex, and we do not
appear to have reached the steady state yet.

\section{Conclusions}
\label{sec:Conclusions}

In conclusion, we have shown that there is a straightforward way to perform
many-body physics calculations in real time for both equilibrium and
nonequilibrium situations by formulating the problem on a Kadanoff-Baym
contour and discretizing it.  There are a number of numerical issues that arise
from this approach, coming from the discretization of the contour and
its truncation, as well as from the discretization of the energy space
needed to perform numerical quadratures.  By using the equilibrium results
as a benchmark, we can see how the different discretization operations
affect the overall accuracy of the calculations.  It seems like the small-time
behavior can be understood fairly well by using reasonable choices for the 
discretizations and the time-domain cutoffs. It is more difficult to obtain
good results for the longer-time behavior.  One also needs to have good
control of the numerics to be able to accurately perform Fourier transformations
to real frequency.  Maintaining control of these different approximations
is made more difficult by the finite computational power that is available to
solve these problems.

The fact that we must iterate our equations to a self-consistent solution
brings a number of unknown issues to the table.  First, we have modified the
Green's function by introducing a time-domain cutoff, which is artificially
changing the boundary conditions.  It is well known that Green's functions are
determined uniquely by their equation of motion and their boundary conditions.
How much of an effect the change of the boundary condition has on the results
is difficult to estimate because of the nonlinearities introduced by the
iterative solution.  Second, we are not able to iterate the equations
for an infinitely long period of time, so the smaller we make the 
discretizations, the fewer iterations we are able to complete. For example,
when the matrices have a size on the order of $700\times 700$, we can
easily perform 50 or more iterations, but we are reduced to about 7
iterations when they are $2000\times 2000$.  It is difficult to tell how
much error is introduced by this.

In the future, we will use this technique to study and analyze better the
behavior of strongly correlated materials in a large electric field.  
Eventually, we hope to be able to generalize this approach to apply it
to multilayered nanostructures and thereby be able to directly calculate
the current-voltage characteristic of a strongly correlated device.

\section{Acknowledgments}
We acknowledge support of the Office of Naval Research under grant
number N00014-99-1-0328 and from the National Science Foundation under
grant number DMR-0210717.  Supercomputer resources  (Cray T3E and X1) were 
provided by
the Arctic Region Supercomputer Center (ARSC) and the Engineering Research
and Development Center (ERDC).

\end{document}

% END of IEEEtest.tex ************